\documentclass{PoS}

\usepackage{amssymb}
\usepackage{amsmath}
\usepackage{graphicx, wrapfig}
\usepackage{booktabs}
\usepackage{color}
\usepackage{textcomp}
\usepackage{multirow}
\usepackage{floatrow}
\usepackage{adjustbox}
\usepackage{setspace}
\usepackage{bm}
\usepackage{caption}
\usepackage{subcaption}
\usepackage{tikz}
\usepackage{longtable}
\usepackage{colortbl}
\usepackage{booktabs}
\usepackage[loadonly]{enumitem}
\usepackage{setspace}

\newfloatcommand{capbtabbox}{table}[][\FBwidth]

\newcommand{\tr}{\mathrm{tr}\,}

\title{Lines of Constant Physics in a Five-Dimensional Gauge-Higgs Unification Scenario}

\ShortTitle{LCP in 5-dimensional Gauge-Higgs Unification}

\author{\speaker{Maurizio Alberti}, Francesco Knechtli  \\ 
      Department of Physics,
      Bergische Universit\"at Wuppertal,\\
      Gaussstr. 20, D-42119 Wuppertal, Germany\\
        E-mail: \email{alberti@uni-wuppertal.de}, \email{knechtli@physik.uni-wuppertal.de}}

\author{Nikos Irges\\
      Department of Physics,
      National Technical University of Athens,\\
      Zografou Campus, GR-15780 Athens, Greece\\
      E-mail: \email{irges@mail.ntua.gr}}

\author{Graham Moir\\
      Department of Applied Mathematics and Theoretical Physics,\\
      Centre for Mathematical Sciences, University of Cambridge,\\
      Wilberforce Road, Cambridge, CB3 0WA, UK\\
      E-mail: \email{graham.moir@damtp.cam.ac.uk}}

\abstract{
We report on the progress in the study of a five-dimensional SU(2) Gauge-Higgs Unification model.
In this non-perturbative study, the Higgs mechanism is triggered by the spontaneous breaking of a
global symmetry. In the same region of the phase diagram, we observe both dimensional reduction
and the ratio of Higgs and Z boson masses to take the value known from experiment.

We present the first results on the construction of a line of constant physics in this region,
including the prediction for the mass scale of the first excited states of the Higgs and gauge bosons.
\\\\\\
\begin{flushright}
\small{
WUB/16-07
}
\end{flushright}

}

\FullConference{34th annual International Symposium on Lattice Field Theory\\
                24-30 July 2016\\
                University of Southampton, UK}

\begin{document}

\section{Introduction}
The discovery of the Higgs field has rendered the Standard Model (SM) complete. However, the
origin of the potential responsible for the spontaneous symmetry breaking (SSB) which is behind the
Brout-Englert-Higgs (BEH) mechanism is still unknown, and the fine tuning necessary to obtain the Higgs
mass measured from experiment raises doubts on the naturalness of the Higgs in the SM.

One possible way to resolve these difficulties is to conjecture the existence of an extra dimension.
From the point of view of our four-dimensional space-time the components of the gauge field in the
fifth dimension would behave as an additional scalar,
giving us a SM-like Higgs field \cite{Manton:1979kb}.
Hosotani \cite{Hosotani:1983} has shown that this idea could work in perturbation theory,
if fermionic degrees of freedom are included in the theory.
Non-perturbatively we have observed the BEH mechanism taking place in a pure gauge theory \cite{IrgesKnechtli:2007,Alberti:2015}
contrarily to perturbation theory's predictions.
For a recent review of GHU studies on the lattice see \cite{KneRin:2016}.

Using an anisotropic action, and enforcing orbifold boundary conditions on the extra dimension,
we have observed that the ratio of the measured H and Z boson masses can reach the experimental value and,
in exactly the same region of the phase diagram, the theory exhibits dimensional reduction \cite{Alberti:2015}.
The simultaneous presence of these two features lends considerable strength to the claim that
reality could contain (at least) one more dimension than what we currently think.\\

The introduction of a fifth dimension, on the other hand, brings a number of new obstacles into the game.
Firstly, such theories are known to be perturbatively non-renormalizable, making it difficult to connect lattice
calculations with possible future experimental results.
Moreover, previous lattice studies have not been able to identify a second order bulk phase transition
point on the phase diagram with periodic boundary conditions along the fifth dimension \cite{Knechtlietal:2011,DelDebb:13,IrgKOut:15}.
Also in our study of the orbifold \cite{Alberti:2015} no second order phase transition could be observed.
Despite this latter drawback, in this work we will show that, at finite cut-off,
the theory is sufficiently well behaved to be used to derive a 4-d effective theory that could reproduce and
possibly complete the gauge-Higgs part of the SM electro-weak sector.
We accomplish this by constructing a line of constant physics (LCP) which shows that,
in the region where $m_H / m_Z > 1$ and dimensional reduction occurs,
several physical quantities remain cut-off independent despite a change of $~ 20 \%$ in the
lattice spacing.

This report is organised as follows. In section \ref{sec:theory} we will construct the
five-dimensional theory on the lattice.
In section \ref{sec:spectrum} we will give a short summary of the previously known results.
In section \ref{sec:LCP} we discuss the construction of a LCP and present our initial results
in this aspect.


\section{Theory}\label{sec:theory}

\begin{figure}[!t]
\center\includegraphics[width=0.4\textwidth]{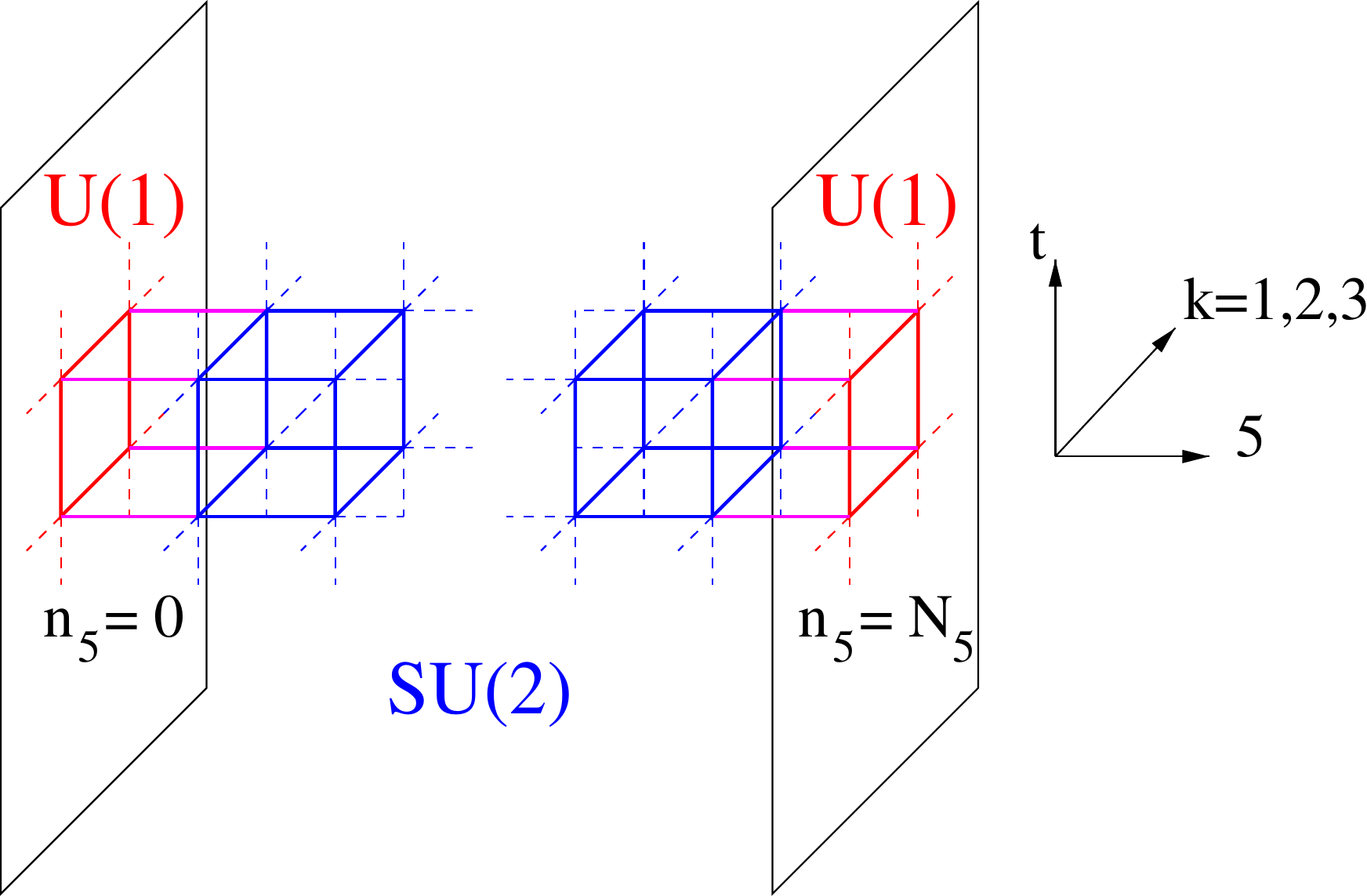}
\caption{ {\footnotesize A sketch of the orbifold lattice. The $SU(2)$ links are depicted in blue
and the $U(1)$ links on the boundary in red. The magenta links connecting the two are so-called
hybrid links, which gauge-transform as $SU(2)$ on one end and as $U(1)$ on the other.}}
\label{fig:orbifold}
\end{figure}

We study a five-dimensional $SU(2)$ gauge theory.
On the lattice, the theory is defined by the action
  \begin{equation}
    S_W^{orb} = \frac{\beta_{4}}{2}\sum_{P_{4}}{w\cdot \tr{\left\{ 1 - P_{4} \right\}}} + \frac{\beta_{5}}{2} \sum_{P_{5}}{\tr{\left\{ 1 - P_{5}\right\}}}
\label{eq:action}
  \end{equation}
where $\beta_{4}$ and $\beta_{5}$ are the gauge couplings associated with Wilson plaquettes spanning
the standard four dimensions ($P_{4}$) and the fifth dimension ($P_{5}$) respectively.
In the sums of eq. \ref{eq:action} plaquettes are counted with one orientation only.
As already mentioned, we apply orbifold boundary conditions in the extra dimension and leave the other
four dimensions periodic\footnote{For a more detailed definition of the theory and its
symmetries on the orbifold, see \cite{Irges:2013rya}.}.
A consequence of this choice of boundary conditions is that, at the orbifold fixed points,
the gauge group is explicitly broken down to $U(1)$.
On the same boundaries, moreover, the weight $w$ which is associated with plaquettes $P_4$
takes a value $w=1/2$, whereas in the orbifold's bulk it remains 1.
It follows that the theory is defined on the interval $I= \left\{n_{\mu}, 0\leq n_5\leq N_5\right\}$, as depicted in figure \ref{fig:orbifold}.
The theory is completely defined by the three parameters $\beta_4$, $\beta_5$ and $N_5$.

On the boundaries of the extra dimension, assuming that some mechanism of dimensional reduction would
take place\footnote{In section \ref{sec:spectrum} we will show that this is indeed the case.}, this
gauge theory is expected to reduce to the Abelian Higgs model.
The scalar degrees of freedom can be probed, for example, by taking the trace of Polyakov loops
winding in the extra dimension, while a vector boson operator can be constructed by displacing them in a
spatial direction \cite{IrgesKnechtli:2007}.


\section{Phase Diagram and Spectrum}\label{sec:spectrum}

This section contains a summarized version of the results from \cite{Alberti:2015}; we report them here in order
to better put the LCP study into context.

\begin{figure}[!t]
\center\includegraphics[width=0.62\textwidth]{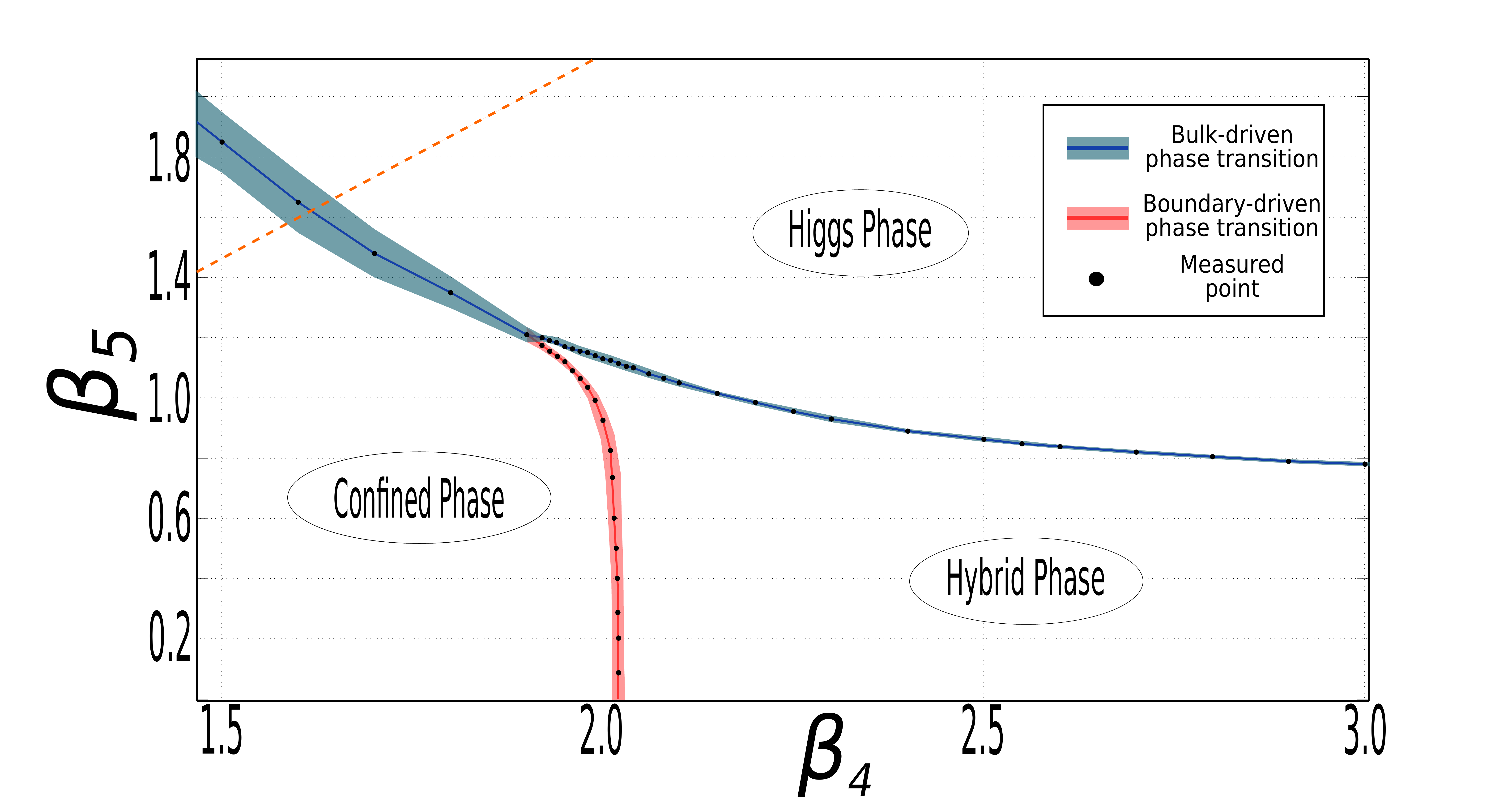}
\caption{ {\footnotesize Phase diagram of the theory.}}
\label{fig:phase-diag}
\end{figure}

As shown in figure \ref{fig:phase-diag}, the theory contains three phases, characterized by the shape of the
static quark potential, which we measure using HYP-smeared Wilson loops \cite{HasKnech:2001}.
All phase transition points in figure \ref{fig:phase-diag} are of first order.
In the \emph{confined} phase the potential is string-like everywhere in the 5-d space-time.

In the \emph{hybrid} phase we continue to observe confined physics in the bulk $SU(2)$ hyperplanes,
while the two $U(1)$ boundaries now are best described by a Coulomb-type potential.
The boundary-driven transition, which occurs at $\beta_4\simeq 2.02$,
is entirely consistent with that observed on a pure four-dimensional $U(1)$ system \cite{Arnold:2003}
and with the \emph{confined-Coulomb} phase transition observed at low Higgs-gauge coupling
in the 4-d Abelian-Higgs model \cite{Jansen:1986,Evertz:87}.
This behaviour shows that in this phase the physical content of a 4d hyperplane is
entirely decided by the gauge group, and therefore
the layers must be very weakly coupled. This result, which at the limit $\beta_5\rightarrow 0$ is obvious,
persists up to the hybrid-Higgs phase transition, with small effects due to $\beta_5 > 0$,
which are visible by measuring the Sommer scale \cite{SScale} in the confined bulk (but not in
the shape of the potential), only in the vicinity of the phase transition.

The third phase is dubbed \emph{Higgs} phase because it is in this region that SSB is observed.
Here, massive scalar and vector boson can be measured and the static potential is, consistently, of Yukawa type.
Moreover, the fit to the potential is confirmed by spectroscopic measurements,
which confirm the presence of a massive gauge boson and reveal that of a massive scalar.
From the functional shape of the static potential one can obtain information on  the dimensionality of the system.
From such observations it appears that the dimensional reduction observed in the hybrid phase persists,
in some respects, also on the other side of the phase transition: when in the vicinity of the phase transition,
the static potential measured on the orbifold's boundary is always a 4-dimensional Yukawa potential,
while in the orbifold's bulk a 5-dimensional Yukawa potential is observed\footnote{Measurements subsequent the publication
 of \cite{Alberti:2015} indicate that, if one measures near enough to the phase transition, the potential can appear of the 4-dimensional
Yukawa type also in the bulk.}. This dimensional reduction mechanism hints at localization on the orbifold boundary.\\

\begin{figure}[!t]
\center\includegraphics[width=0.6\textwidth]{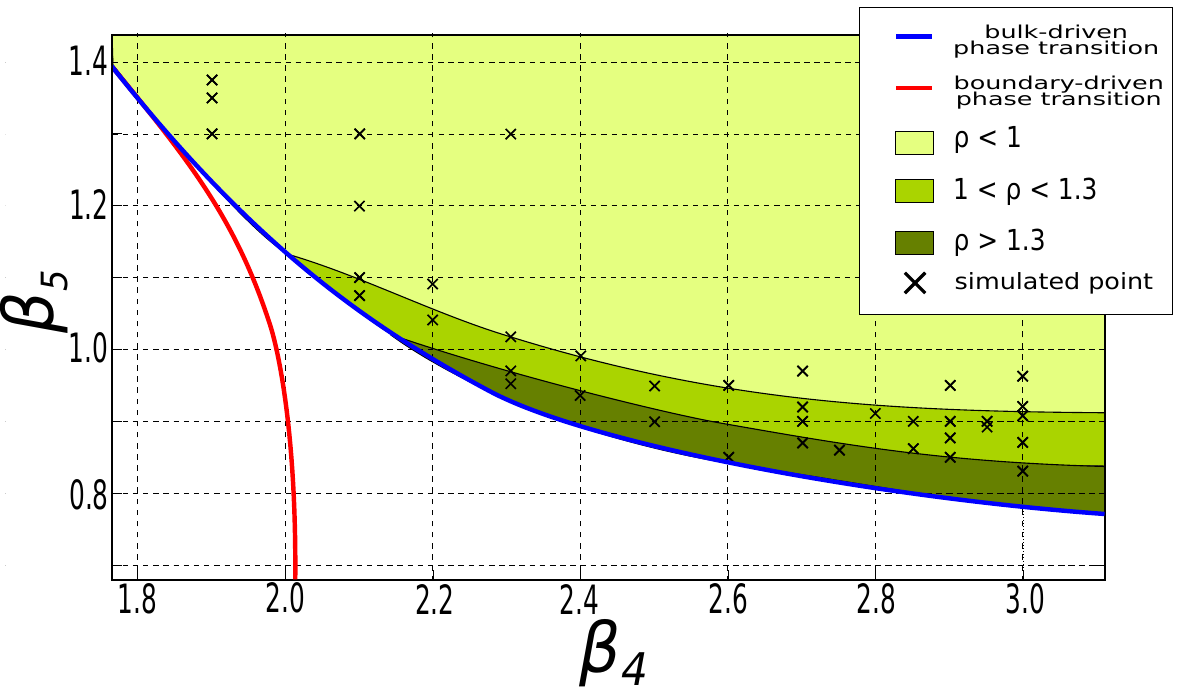}
\caption{ {\footnotesize Contour plot of $\rho\equiv m_H / m_Z$ varying the theory's $\beta_4$ and $\beta_5$ parameters
            The two darker shades of green indicate the region where the boson masses exhibit the correct hierarchy.
            All the measurements in this plots were performed at $N_5 = 4$.}}
\label{fig:contour-rho}
\end{figure}

A very interesting result, and a very important one with respect to the plausibility of this GHU theory, is that,
in the same region of the Higgs phase where dimensional reduction is observed, the ratio $\rho \equiv m_H / m_Z$
of the Higgs to the scalar mass reaches the value measured from experiment. Figure \ref{fig:contour-rho} shows
a detail of the phase diagram, where in the Higgs phase different shades of green highlight regions with a different
$\rho$ regime. In the regions represented by the two darker shades of green that both dimensional reduction and
a correct mass hierarchy have been simultaneously observed.


\section{Lines of Constant Physics}\label{sec:LCP}
\vspace{-0.3cm}
\subsection{LCPs in Theory}\label{ssec:lcp-thepry}
When studying lattice theories which possess a continuum limit, lines of constant physics are employed to extrapolate the
"physical" value of any measurable quantity. This is done by constructing a line in the parameter space,
on which a certain number of dimensionless physical quantities are kept fixed, usually at their experimental value.
In a theory with $N$ parameters only $N-1$ dimensionless quantities can be kept fixed on any such line; the value of all remaining
observables might instead show a dependency on the cut-off.
The power of the method resides in the fact that, when the LCP approaches the continuum limit, \textbf{all} quantities
reach values that are free from cut-off effects. In other words, if the $N-1$ quantities which define the LCP are set
to their physical value, at the critical point all quantities should reach their physical value as well.

To simplify the process described above, one can even fix only $M<N-1$ quantities to construct a line of \emph{partially} constant
physics (LPCP), which presents the same behaviour when approaching the continuum limit, but which will then not be unique \cite{MontMunst}.\\

In a theory without a continuum limit, or with only a trivial one, the construction of a line of constant physics
will serve a different purpose, i.e. to verify whether or not the theory can be used as an effective theory
which is valid for energies much lower than the finite cut-off.

In order to do so, one would construct an LCP exactly as in a theory with a continuum limit: keeping $N-1$ dimensionless quantities fixed
and changing the parameters enough to observe a sizeable change in the cut-off.
As stated above, there are no guarantees that quantities other than the $N-1$ fixed ones will not
change on this line; they are, in fact, expected to vary due to finite lattice spacing effects.
Nevertheless, one might find that there exists a portion of the LCP where all observed quantities appear, within errors, to be constant.
This is equivalent to the statement that the size of the cut-off effects is smaller than the statistical accuracy on these observables.

On such a segment of the LCP the theory, even if it, as in our case, might be non-renormalizable, can be used as an effective theory to
describe reality. Moreover, if a way to set the scale is known, the maximal energy for which the effective theory is valid
can be computed from the minimal lattice spacing on the segment.\\

Since our goal is to test the possibility of using the 5-dimensional theory on the orbifold to reproduce at low energies
the EW sector of the SM, we concentrate on the construction of LCPs in the theory's Higgs phase.
As a first step, to decrease the technical challenge of exploring a 3-dimensional parameter space, we will attempt to construct
lines of partially constant physics at a fixed value of $N_5$, by imposing only the requirement that $\rho \equiv m_H / m_Z = const$.
In order to parametrise the change in the lattice spacing we will observe the change in the quantity $a_4 m_Z$.
We will then repeat the procedure at a few different values of $N_5$; if a full LCP can be found
it will be lying on some of the points on the LPCPs constructed at fixed $N_5$.

\subsection{LCP Observation}
\begin{figure}[t!]
    \includegraphics[width=0.85\linewidth]{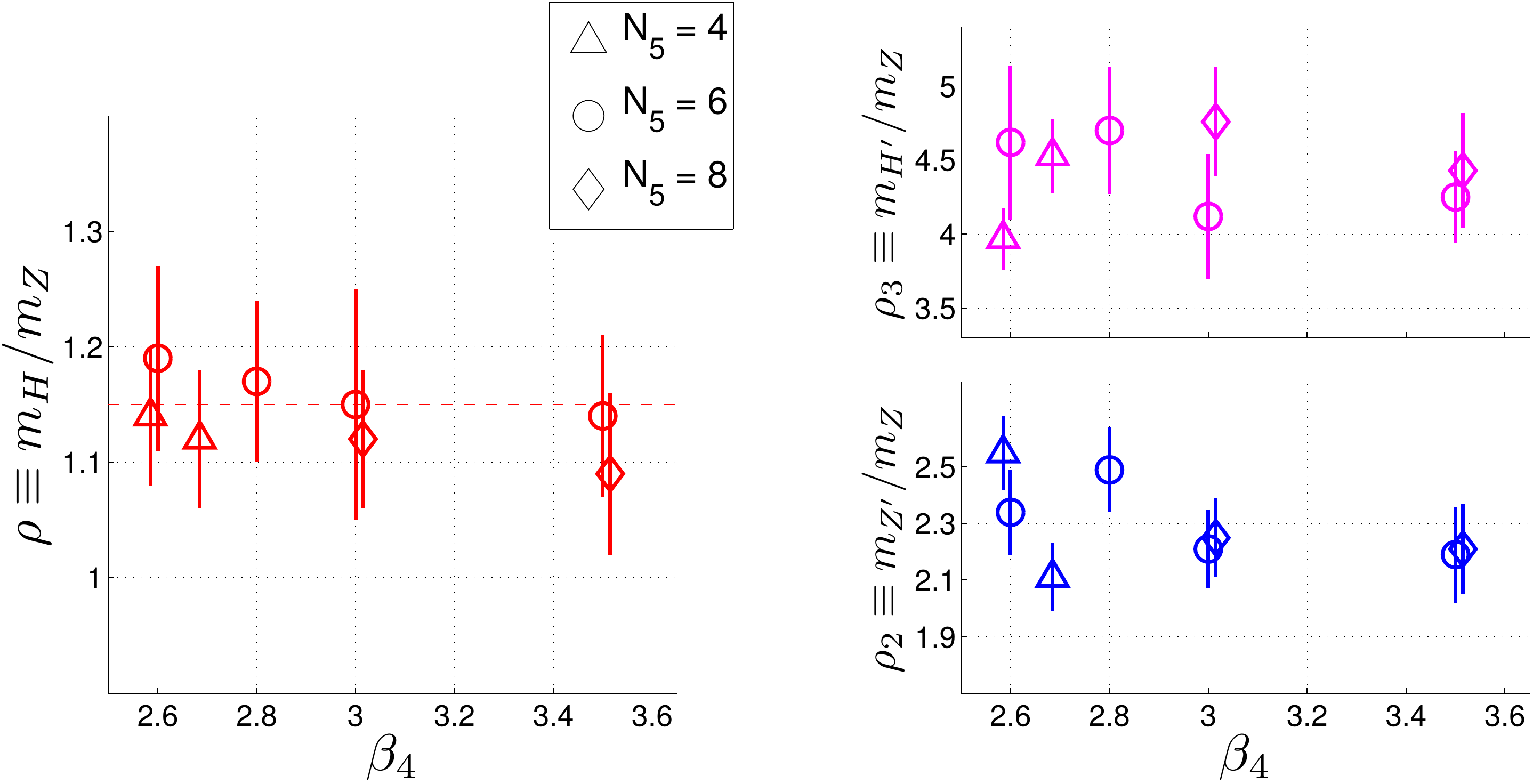}
    \caption{ {\footnotesize The left panel shows the values of $\rho$ on the line of partially constant physics defined by $\rho=1.15$.
                In the right panel the control quantities $\rho_2$ and $\rho_3$, measured at the same parameter values, are shown.}}
    \label{fig:LCP-rho}
\end{figure}

We search for lines of partially constant physics by imposing the requirement $\rho \equiv m_H/m_Z = 1.15$ within statistical accuracy.
Moreover, we measure the other two dimensionless quantities $\rho_2\equiv m_{Z'}/m_Z$ and $\rho_3\equiv m_{H'}/m_Z$ to check whether
the points found on the LPCPs might belong to a more restricting LCP, e.g. if we observe that $\rho_2$
remains constant w.r.t. the change of lattice spacing,
or even better that the model can be used as an effective theory in the region we search,
if all three quantities should show no cut-off dependence.
We have measured these quantities at $N_5 = 4,6,8$. Figure \ref{fig:LCP-rho} shows the first promising results in this direction.
The left panel of the figure shows the values of $\rho$ as a function of the lattice coupling $\beta_4$; the value of the second coupling $\beta_5$
has been adjusted to remain on the LPCP defined by $\rho=1.15$ with a precision of $\Delta \rho = 0.05$.
while it appears to be more sensitive to the changes in $N_5$. Considering all measurements, the quantity $a_4 m_Z$ has changed of $~20\%$, going
from $0.155$ measured at $N_5=4$ to $0.131$ measured at $N_5=8$.

The right panel of figure \ref{fig:LCP-rho} shows that $\rho_2$ and $\rho_3$ remain fairly constant on the LCPC,
which suggests that cut-off effects are contained.
It is worth noting that we measure a sizeable gap between the ground and the excited states,
with $m_{Z'}$ being $~2.3$ and $m_{H'} ~4.5$ times larger than $m_Z$.

The choice of $\rho=1.15$ instead of the physical value of $1.39$ is only due to the lesser difficulty in performing Monte Carlo
simulations slightly further away from the phase transition\footnote{As shown in \cite{Alberti:2015}, the value of $\rho$ increases as
the parameters approach the Higgs-hybrid phase transition.}. Since this study is a proof-of-concept for the viability of the model
rather than a study of its physical properties, the existence of any line of constant physics
in the region of the phase diagram in which the system is dimensionally reduced is the most important result.
Nevertheless, work on constructing an LCP at physical $\rho\simeq 1.39$ is underway.


\section{Conclusion}\label{sec:conclusion}
In the previous sections we have shown how this 5-dimensional GHU theory contains all the features which would be required in order for it
to be a viable alternative, not affected by the hierarchy problem, to the current SM description of the electroweak sector.
In older works it has been shown that, in a definite region of the Higgs phase, a mass hierarchy resembling that observed in
experiments can be achieved, and that in the same region the system appears dimensionally reduced.

In the present addition to the study, we showed that these two absolutely non-trivial features are not just a result
of some particularly lucky combination of parameters, but they persist despite a change of about 20\% in the lattice spacing.
Although this value is not yet sufficient to allow a conclusive statement on the matter,
the results presented here point in the direction of this theory being worth studying further as a possible effective theory
at energies below the cut-off.

\subsection*{Acknowledgements}
{\small
This work has been supported by the Deutsche Forschungsgemeinschaft (DFG) under contract KN 947/1-2.
The authors gratefully acknowledge the computing time granted by the John von Neumann Institute for
Computing (NIC) and provided on the supercomputer JURECA at J{\"u}lich Supercomputing Centre (JSC).
G.M. also acknowledges support from the Herchel Smith fund at the University of Cambridge.}

\end{document}